\documentclass[reprint,amsmath,amssymb,aps,pra]{revtex4-2}

\usepackage[english]{babel}
\usepackage{graphicx}% Include figure files
\usepackage{dcolumn}% Align table columns on decimal point
\usepackage{bm}% bold math
\date{Received 24 January 2023. Published: 12 may 2023}

\begin{document}

% \preprint{APS/123-QED}

\title{High-resolution dynamic consistency analysis of \\ photonic time-delay reservoir computer}% 

\author{Lucas Oliverio $^{*,1,2}$}
\author{Damien Rontani $^{1,2}$}
\author{Marc Sciamanna $^{1,2}$}
\affiliation{
Chaire Photonique, LMOPS, CentraleSupélec, 57070 Metz, France\\
Université de Lorraine, CentraleSupélec, LMOPS, 57070 Metz, France\\
$^*$ Corresponding author : lucas.oliverio@centralesupelec.fr
}

\begin{abstract}
We numerically investigate a time-delayed reservoir computer architecture based on a single mode laser diode with optical injection and optical feedback.
Through a high-resolution parametric analysis, we reveal unforeseen regions of high dynamical consistency. 
We demonstrate furthermore that the best computing performance is not achieved at the edge of consistency as previously suggested in a coarser parametric analysis.
This region of high consistency and optimal reservoir performances are highly sensitive to the data input modulation format.

\end{abstract}

\maketitle
\section{Introduction}
The rapid increase of information exchange in optical networks requires novel processing paradigms. It has been made possible by the development of machine learning and in particular by photonic-based artificial neural network architectures.
Reservoir computing (RC) is a machine learning technique that trains artificial neural networks made of recurrent connections by regression on the output layer \cite{jaeger_harnessing_2004}.
Several reservoir computer architectures have been proposed using coupled photonic emitters \cite{brunner_photonic_2017}, silicon photonic chip \cite{vandoorne_experimental_2014} or free-space systems \cite{antonik_human_2019}. 
However, having a large number of nodes in a RC is still relatively challenging experimentally. 
To overcome this difficulty, the suggestion has been made to use a single physical node but with a delayed feedback loop, resulting in a so called time-delayed reservoir computer (TDRC) \cite{appeltant_information_2011}. 
TDRC have been proposed using optoelectronic systems \cite{martinenghi_photonic_2012,soriano_optoelectronic_2013,argyris_comparison_2020}, nonlinear passive devices \cite{duport_all-optical_2012,sunada_photonic_2019}, semiconductor lasers with optoelectronic \cite{paquot_optoelectronic_2012,larger_photonic_2012} or optical feedback \cite{vatin_experimental_2019,brunner_parallel_2013,nguimdo_fast_2014,hicke_information_2013,takano_compact_2018, owen-newns_ghz_2022}.

Parametric analysis of the RC performances has been carried out often with a limited parametric resolution, both theoretically or experimentally \cite{nakayama_laser_2016,kanno_reservoir_2022,bauwens_influence_2022,bueno_conditions_2017,oliver_consistency_2015,estebanez_accelerating_2020}.
These first parametric analyses help determining the optimal operating point of the RC, but they typically do not address the question of how the performance relates to the dynamics of the laser.

In this paper we conduct an in-depth high resolution study of both the metrics linked to the RC performances and the system dynamical consistency.
Consistency measures the ability to generate similar responses for similar inputs \cite{uchida_consistency_2004}. In an optical injection system, the parameter region leading to high consistency corresponds to the injection locking area with a smooth and continuous boundary \cite{kanno_consistency_2012}. However, we show here that the boundary of consistency region is rather structured with a series of tongues that repeat periodically and in which high consistency is still measured. The boundary is also dramatically impacted by the input data modulation. Finally, it is known that the consistency of the system is directly linked to the performance of the system. In this framework we study the impact of these novel consistency regions on classical benchmark tasks for RC.
Our theoretical analysis brings insight into recent experiments \cite{bueno_conditions_2017}, but also uncovers new parametric regions of high consistency and good performances that have not been observed so far.

\section{Numerical Model}
We simulate a TDRC based on a semiconductor laser. We consider a single mode Lang-Kobayashi model \cite{lang_external_1980} extended to account for optical injection and optical feedback:

\begin{align}
    \label{eq:model1}
    \frac{dE(t)}{dt}\: =\:\: &\frac{(1+i\alpha)}{2}  \left[G\left[N-N_0\right]-\frac{1}{\tau_{ph}} \right]E(t) \\
    &+\kappa_f E(t-\tau) e^{i\omega\tau}+ \kappa_{inj}E_{inj}(t)e^{-i\Delta\omega t},  \notag \\
    % \\
    \frac{dN(t)}{dt}\:=\:\: &p\,J_{th}-\frac{N}{\tau_s}-G\left[N-N_0\right] |E(t)|^2
    \label{eq:model2} 
\end{align}

where, $E(t)$ is the slow varying amplitude of the complex electric field and $N(t)$ is the carrier density. $\alpha$ is the linewidth enhancement factor. $\tau_{ph}$ is the photon lifetime. $\tau_s$ is the carrier lifetime. $G$ is the gain coefficient. $N_0$ is the carrier density at transparency. $\kappa_f$ is the feedback strength. $\tau$ is the length of the delay line. $\kappa_{inj}$ is the injection strength. $p$ is the normalised injection current ; $J_{th}$ is the threshold current. $\Delta\omega$ is the angular frequency detuning ($\Delta\omega=2\pi\Delta f=2\pi(f_d-f_r)$) between the drive  and the response laser which, set to zero. 
$E_{inj}(t)$ is the injected field optically modulated by a Mach-Zehnder modulator (MZM) as follows: $E_{inj}(t)=\sqrt{P_{inj}} \, /2 \left\{ 1+\textrm{exp} \left( i\left[V(t)/V_{\pi}+\phi_{DC} \right] \right)\right\}$ with $V(t)=\gamma \: D(t)$. Where, $P_{inj}$ is the amplitude of the injected field, $\gamma$ is the scaling factor, $V_\pi$ is the voltage resulting in a $\pi$ phase-shift between the arms of the modulator, $\phi_{DC}$ is the bias of the modulator and $D(t)$ is the input signal in $\left[-1;+1\right]$.
We simulate the model with a second-order Runge-Kutta method on a super computer for high-resolution analysis. We fix $\alpha=5$, $\tau_{ph}=1.927 \times 10^{-12}$ s, $\tau_s=2.04 \times 10^{-9}$ s, $G=8.4 \times 10^{-13}\,\textrm{m}^3 \textrm{s}^{-1}$, $N_0=1.40 \times 10^{24}\,\textrm{m}^{-3}$, $\tau=1$ ns, $p=1.015$ and $J_{th}=9.892\times10^{32}\,\textrm{m}^{-3}\textrm{s}^{-1}$.

%%%%%%%%%%%%%%%%%%%%%%%%%%%%%%%%%%%%%%%%%%%%%%%%%%%%%%%%%%%%%%%%

\begin{figure}[ht]
    \centering
    \includegraphics[width=\linewidth]{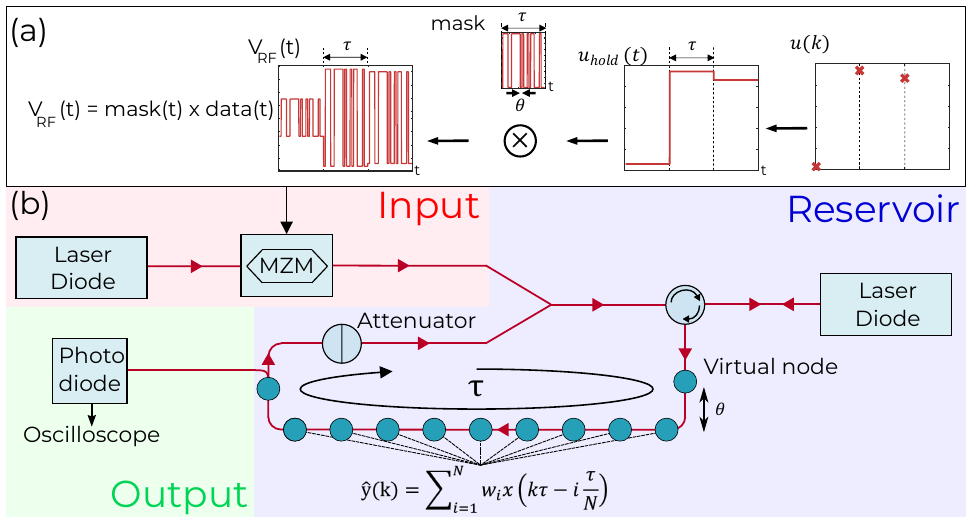}
    \caption{(a) Masking procedure. The input u(k) is convoluted with a $\tau$-length mask of 50 random values in $\{-1;1\}$. (b) The information is then optically injected into the reservoir using a Mach-Zehnder modulator. The light goes into the non-linear node, a laser diode. The feedback is implemented with a delay loop that can be implemented through fiber optics or free space, feedback strength is adjusted with an attenuator. The output layer corresponds to the reading of the reservoir states sampled at time $\theta$.}
    \label{fig:figure_1}
\end{figure}

Figure \ref{fig:figure_1} shows the system diagram, a TDRC based on a laser diode with an optical feedback of time delay $\tau$. The feedback loop can be implemented through fiber optics or free space. We choose the inter-neuron interval $\theta$ as one-fifth of the typical response time of the non-linear node \cite{appeltant_information_2011}.
The node interval is $\theta=20$ ps and the delay $\tau=1$ ns, thus there are $ N=\tau / \theta= \: 50$ neurons.
The information is optically injected into the reservoir, using a Mach-Zehnder modulator. 
As in \cite{appeltant_constructing_2015}, the signal is first multiplied by a binary $\tau-$long mask composed of a piecewise function of $N$ values randomly chosen in $\{-1;+1\}$.
The output layer corresponds to the reading of the reservoir states sampled at time $\theta$.
The training phase consists of finding the optimal weights by minimising the error between the output and the target using Moore-Penrose inversion.

%%%%%%%%%%%%%%%%%%%%%%%%%%%%%%%%%%%%%%%%%%%%%%%%%%%%%%%%%%%%%%%%

As introduced in \cite{uchida_consistency_2004}, the consistency represents the ability of a dynamical system to respond in the same way when it is perturbed by the same external signal starting from slightly different initial conditions. Consistency is the correlation between two output signals, $C = \langle (I_1-\overline{I_1})(I_2-\overline{I_2})\rangle/\sigma_1\sigma_2$, where $I_i$ is the intensity, $\overline{I_i}$ is the average intensity and $\sigma_i$ is the standard deviation of $i^{th}$ response.
To be consistent is a necessary condition for RC \cite{oliver_consistency_2015,uchida_consistency_2004,kanno_consistency_2012,nakayama_laser_2016,bueno_conditions_2017}.
\begin{figure}[ht]
    \centering
    \includegraphics[width=\linewidth]{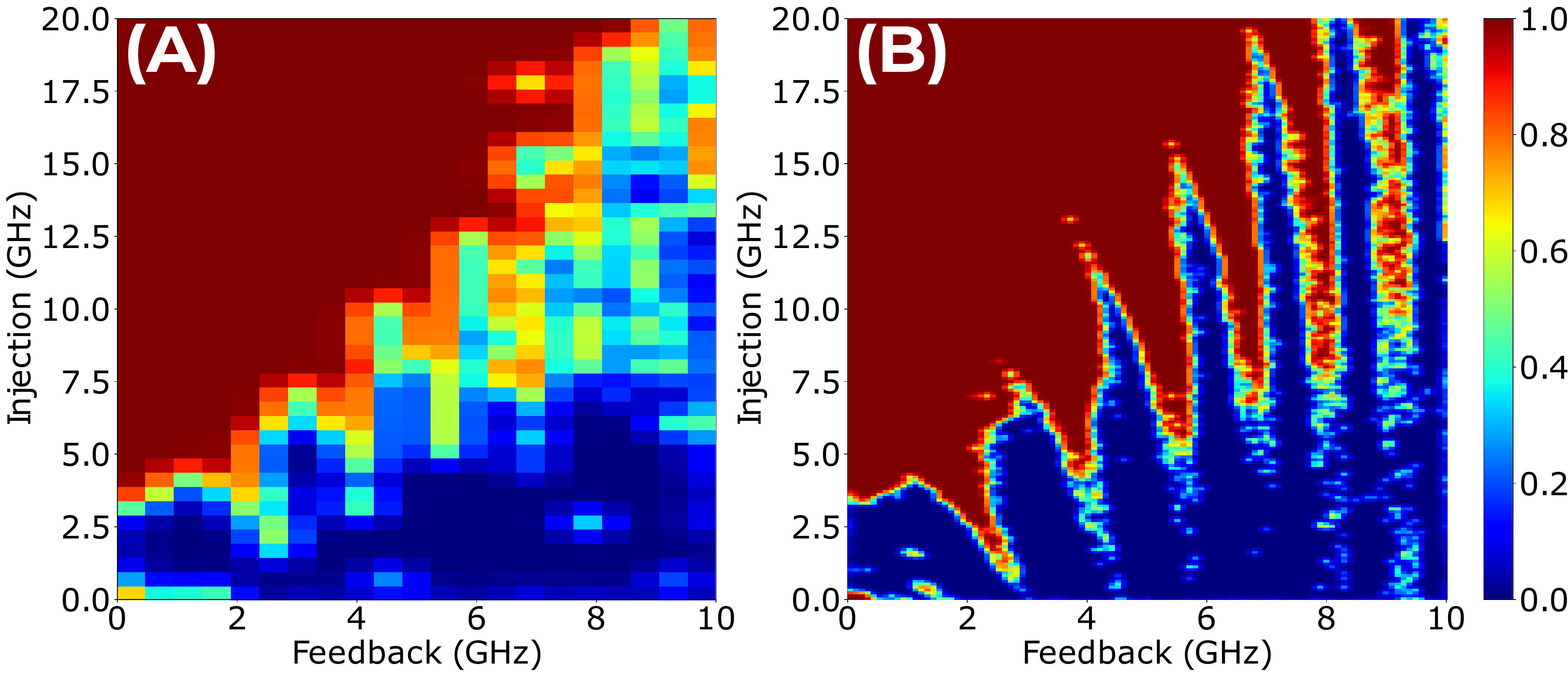}
    \caption{Consistency for $I=1.015\:I_{th}$, $\gamma=\phi_{DC}= \frac{\pi}{4} $ at a resolution step of $\Delta\kappa_{inj}=\Delta\kappa_{f}=$ (a) $0.5$ GHz and (b) $0.1$ GHz.}
    \label{fig:figure_2}
\end{figure}
In our simulations, we have worked in a range of $[0-10]$ GHz feedback and $[0-20]$ GHz injection which corresponds in our system to $[0-8.5]$ \% of feedback and $[0-17]$ \% of injection compared to stationary laser power. This is estimated with $\kappa(\%)=\kappa($GHz$)\:\tau_L$, where $\tau_L=8.5\times10^{-12}$ s is the internal roundtrip time.

\section{Results}
\subsection{Consistency: resolution of the analysis}

We scan two key parameters, the injection strength $\kappa_{inj}$ and the feedback strength $\kappa_{f}$ with a pumping current of $1.015\;I_{th}$.
In Figure \ref{fig:figure_2}(a) at a resolution step of $\Delta\kappa_{inj}=\Delta\kappa_{f}=0.5$ GHz, we observe a smooth boundary of the consistency region. This is reminiscent of consistency region usually observed in previous studies \cite{bueno_conditions_2017, nakayama_laser_2016}.
In Figure \ref{fig:figure_2}(b), we use a finer step size of $\Delta\kappa_{inj}=\Delta\kappa_{f}=0.1$ GHz. Such a small step corresponds to a variation of about $10$ $\mu$W of injected or retro-injected power for a laser emitting $10$ mW. Such a small variation is easily achievable with the current fine tuning of a variable optical attenuator. 
Such a fine analysis reveals the existence of a \textit{tongue}-shaped geometry of the consistency boundary.
Figure \ref{fig:figure_2} therefore shows that a transition to consistency is not smooth and continuous but rather abrupt and discrete, hence revisiting some earlier statements \cite{kanno_consistency_2012,bueno_conditions_2017}.

\begin{figure}
    \centering
    \includegraphics[width=\linewidth]{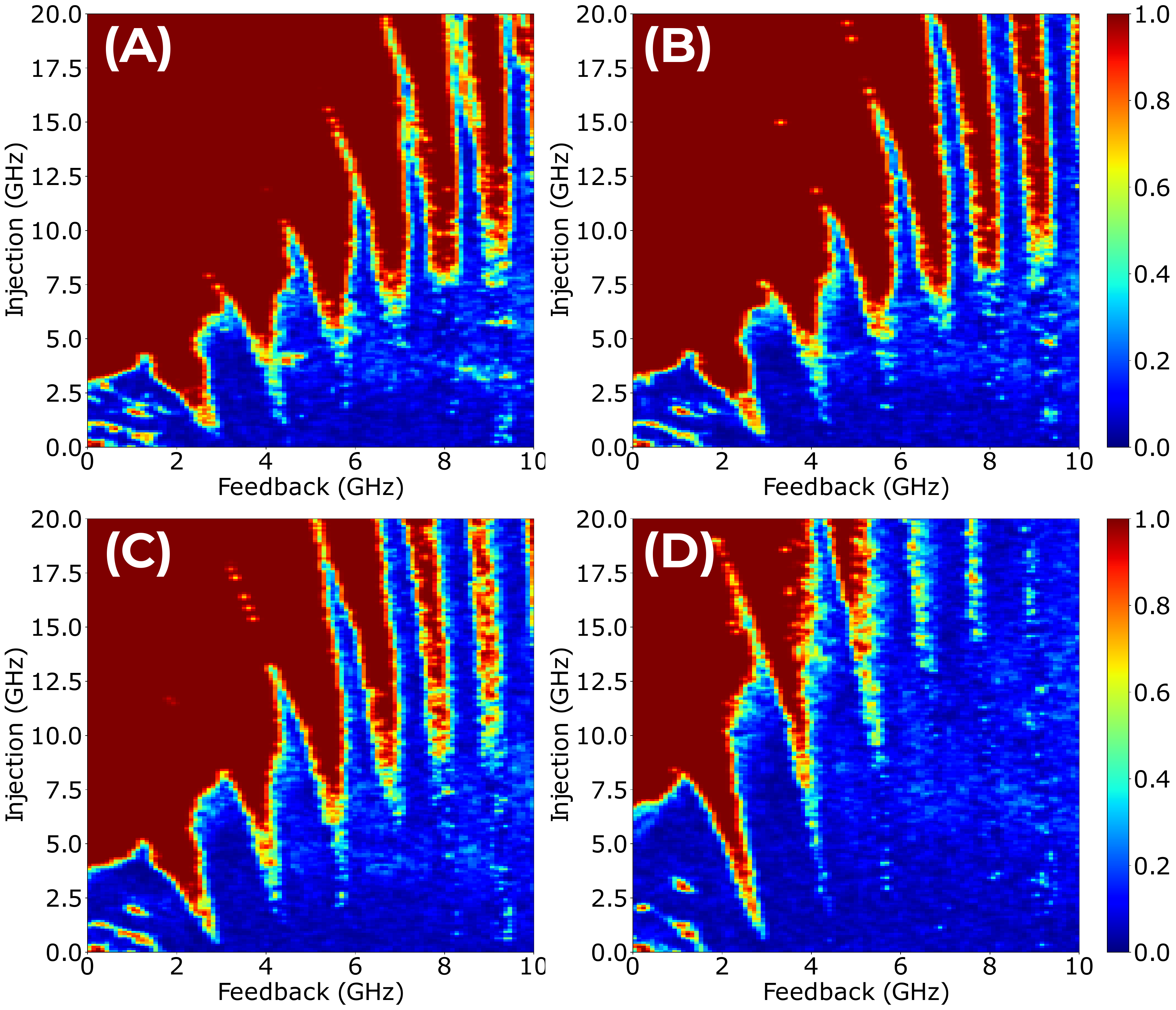}
    \caption{Consistency at a resolution step of $\Delta\kappa_{inj}=\Delta\kappa_{f}=0.1$ GHz for $I=1.015\:I_{th}$, $\phi_{DC}=0$ and $\gamma=$ (a) $\frac{\pi}{8}$, (b) $\frac{\pi}{4}$, (c) $\frac{\pi}{2}$, (d) $\pi$.}
    \label{fig:figure_3}
\end{figure}

\subsection{Impact of the modulation depth}
We now analyse the impact of the input modulation. The  MZM is controlled by a voltage that ranges from $[-\pi V_{\pi} ;+\pi V_{\pi} ]$ what can be modified by $\gamma$ the scaling factor and $\phi_{DC}$ the bias voltage, which are two experimentally tunable parameters. 
Figure \ref{fig:figure_3} shows consistency maps with $\gamma\,\in\{\frac{\pi}{8},\: \frac{\pi}{4},\: \frac{\pi}{2},\: \pi \}$ and  $\phi_{DC}=0$ values, we study the impact of $\gamma$ and $\phi_{DC}$ values. 
We see that, as $\gamma$ increases, the consistency region shrinks toward larger $\kappa_{inj}$. When $\phi_{DC}$ increases, the consistency region shifts globally upwards (stronger injections). By having also considered $\phi_{DC}\,\in\{0,\: \frac{\pi}{4},\: \frac{\pi}{2},\: \frac{3\pi}{4},\: \pi\}$, we observe the same dynamic, not shown here.  
The consistency of the system shows therefore a high dependence on the modulation depth. These experimentally tunable parameters modify the size and location of the region of high consistency significantly.
\begin{figure}[ht]
    \centering
    \includegraphics[width=\linewidth]{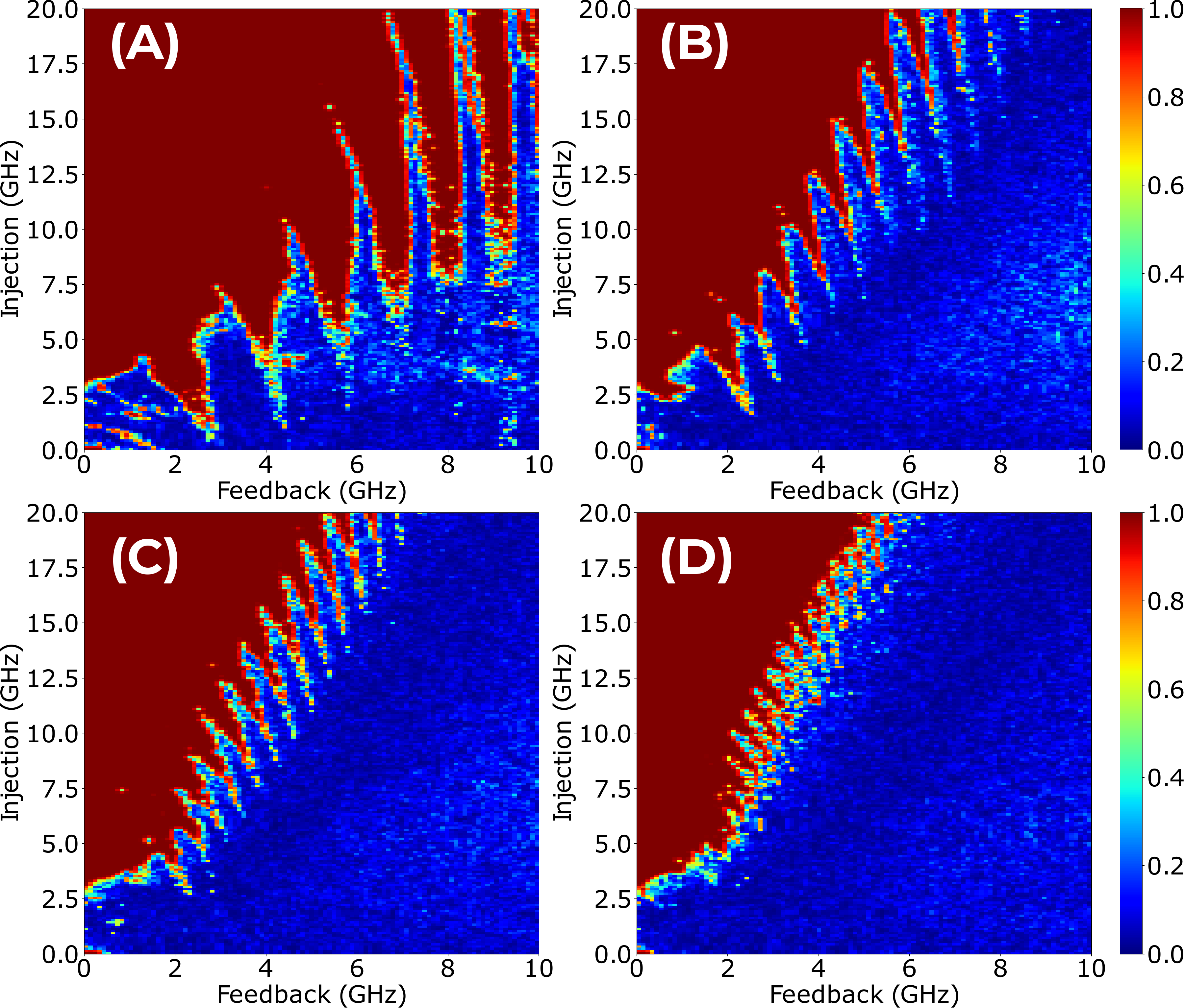}
    \caption{Consistency at a resolution step of $\Delta\kappa_{inj}=\Delta\kappa_{f}=0.1$ GHz for $I=1.015\:I_{th}$, $\gamma=\frac{\pi}{8}$ and $\phi_{DC}=0$ considering different values of delay. $\tau=$ (a) 1 ns , (b) 2 ns, (c) 3 ns, (d) 5 ns.
    }
    \label{fig:figure_4}
\end{figure}

\subsection{Impact of delay}
In Figure \ref{fig:figure_4} we make consistency maps with $\gamma=\frac{\pi}{8}$ and $\phi_{DC}=0$ values and we study the impact of the value of the delay time $\tau$ on the consistency region, at a resolution step of $\Delta\kappa_{inj}=\Delta\kappa_{f}=0.1$ GHz. We consider $\tau\textrm{ in } \{1;2;3;5\}$ ns.
We observe the significant impact of the delay on the topology of the consistency area at the immediate vicinity of the tongues: both the size and the repetition rate of the tongues is modified by the value of the time delay. The region of high consistency away from the border is not visibly affected by these delay changes, and is preserved overall as we can see on Figure \ref{fig:figure_4}.
By measuring the $\Delta\kappa_{f}$ separating two consecutive tongues, we not only confirm that this spacing remains equal from tongue to tongue, whatever the delay, but also that this spacing is inversely proportional to the delay. 
We note that the transition from an abrupt and discontinuous boundary is related to the values of delays being large compared to the laser, in particular $\tau >\tau_{RO}$, where $\tau_{RO}$ is the frequency of relaxation oscillations of the laser. At our operating point, $\tau_{RO}\simeq$ 2.2 ns.
We also check the impact of the pump current on the consistency map. We notice that an increase of the current greatly degrades the consistency, by shrinking the consistency region. 
Thus the system becomes less likely to operate under favorable conditions, hence justifying why almost all reported experimental studies show better performances near threshold \cite{hicke_information_2013}.

\subsection{Performance analysis}
\begin{figure}[ht]
    \centering
    \includegraphics[width=\linewidth]{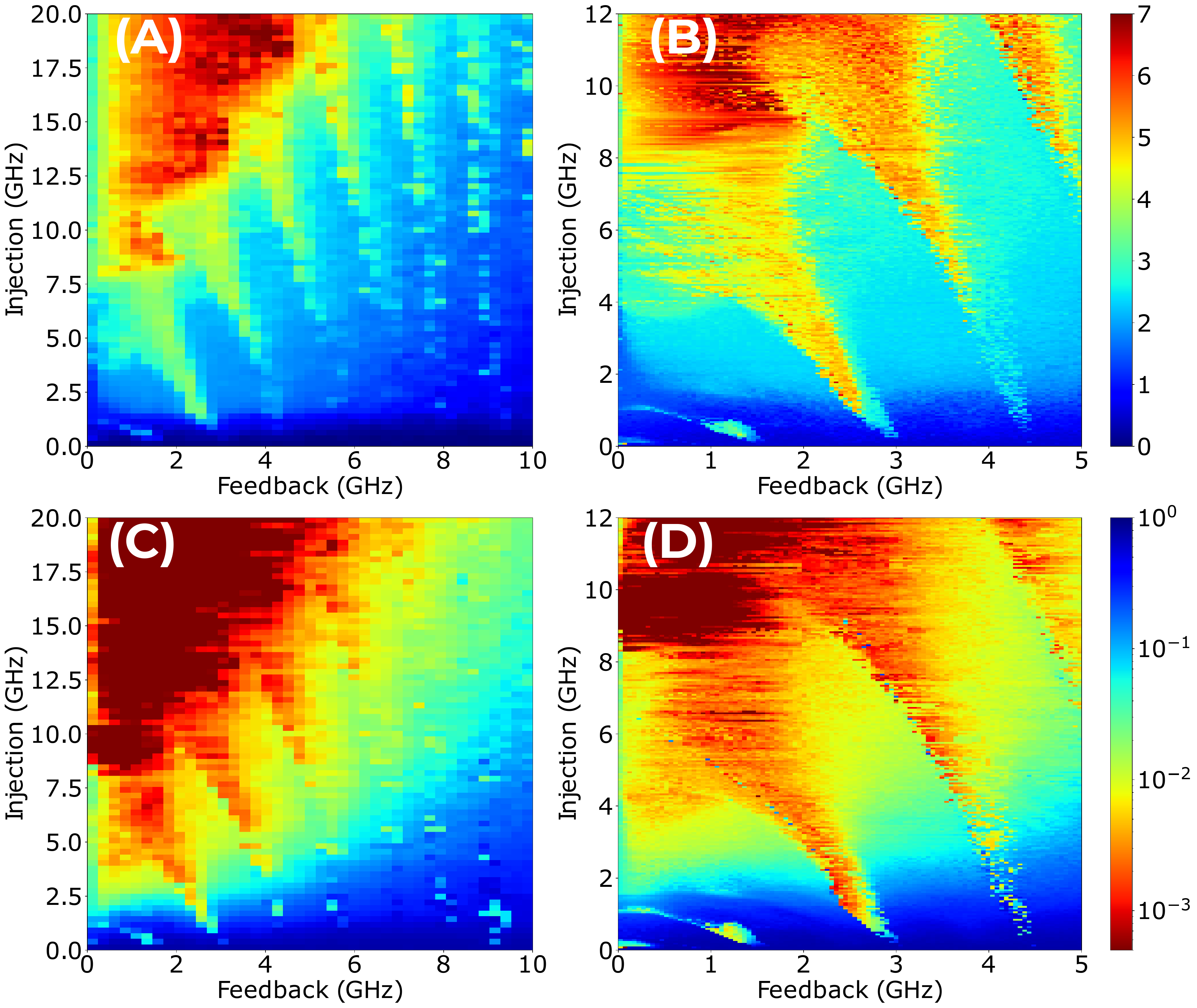}
    \caption{Memory capacity at a resolution step of $\Delta\kappa_{f}=\Delta\kappa_{inj}=$ 
    (a) $0.25$ GHz,
    (b) $0.05$ GHz.
    BER for the nonlinear channel equalisation task at a resolution step of $\Delta\kappa_{f}=\Delta\kappa_{inj}=$  
    (c) $0.25$ GHz, 
    (d) $0.05$ GHz. 
    For $I=1.015\:I_{th}$, $\phi_{DC}=0$ and $\gamma=\frac{\pi}{8}$
    }
    \label{fig:figure_5}
\end{figure}

Let us now study how the consistency analysis relates to the RC performances. We first investigate the linear memory capacity (MC) which reflects the number of steps in the past that the system can remember \cite{jaeger_tutorial_2002} and is defined as follows: $\mu_c=\sum_{i=0}^{\infty}corr\left[ y_i(k),y(k-i)\right]$
where, $y_i(k)$ is a reconstruction of $y(k-i)$, meaning that the RC is trained to reconstruct the $i^{th}$ previous input. The MC cannot be greater than the number of virtual nodes.
The maps in Figure \ref{fig:figure_5} compare the memory capacity of the TDRC with a resolution of $\Delta\kappa_{f}=\Delta\kappa_{inj}=0.25 $ GHz and $\Delta\kappa_{f}=\Delta\kappa_{inj}=0.05 $ GHz, thus we compare the memory values according to the simulation parametric resolution.
In Figure \ref{fig:figure_5}(a) at a resolution step of $\Delta\kappa_{f}=\Delta\kappa_{inj}=0.25 $ GHz, we observe a region with high memory values having a tongue-shaped boundary. They are in line with the consistency tongues obtained under the same conditions in Figure \ref{fig:figure_3}(a), i.e. with $\gamma=\frac{\pi}{8}$ and $\phi_{DC}=0$. The highest values are obtained in the center of the region. By varying the values of $\gamma$ and $\phi_{DC}$, we see the same behaviours as for consistency. Increasing $\gamma$ shrinks the high-memory region towards the larger values of $\kappa_{inj}$ and increasing in $\phi_{DC}$ shifts globally the high-memory region.
In Figure \ref{fig:figure_5}(b), we plot the memory at a higher resolution step of $\Delta\kappa_{f}=\Delta\kappa_{inj}=0.05 $ GHz, focusing on the first tongues. The memory globally degrades as we enter the tongues compared to the core region. Yet, we see that there are regions in the tongues with memory values comparable to the core, hence we show that it is possible for the system to perform well in a tongue as well as in the central region but in a very restricted area.
In Figure \ref{fig:figure_2} we showed that the structuring of the consistency boundary is limited by the resolution. A comparable result is obtained for the memory since with the first resolution the memory seemed low in the tongues. However, we realize that this is not the case with the higher resolution. We also notice that the distribution of the high memory areas in the tongues is not symmetrical.

The previous results suggest that there is a structuring of performance in the parametric plane.
Further proof is provided by exposing our reservoir to a classical task.
For this we have chosen nonlinear channel equalisation which has important practical implications in telecommunication networks \cite{mathews_adaptive_1994}. 
The objective is to reconstruct from observed $u(n)$ the original signal $d(n)$ after it has spread across a non-linear channel.
We consider an input $d(n)$ as an independent and identically distributed random sequence among the values $\{-3;-1;+1;+3\}$.
This signal is transmitted through a nonlinear channel modeled as a system mixing consecutive inputs followed by a noisy memoryless nonlinear system defined as:

\begin{align}
q(n)&=0.08d(n+2)-0.12d(n+1)+d(n)+0.18d(n-1)\notag\\
    &-0.1d(n-2)+0.091d(n-3)-0.05d(n-4)\\
    &+0.04d(n-5)+0.03d(n-6)+0.01d(n-7),\notag
\end{align}
then, $u(n)=q(n)+0.036q(n)^2-0.011q(n)^3+v(n),$

where, $v(n)$ is Gaussian white noise with a given SNR.
The performance on this task is measured with the bit error rate (BER), which is the number of incorrectly inferred bits over the total number of bits reconstructed.
Given that we have a 1 ns delay line, this allows us a processing speed of 1 GSymbols/s.
Figure \ref{fig:figure_5}(c) shows a BER map in the nonlinear channel equalisation task at a resolution step of $\Delta\kappa_{f}=\Delta\kappa_{inj}=0.25 $ GHz and in (d) at a resolution step of $\Delta\kappa_{f}=\Delta\kappa_{inj}=0.05 $ GHz. 
We use 15 000 symbols, 3 000 for the training and 12 000 for the testing.
The \textit{tongue}-shaped structure is again found and overlaps with that of consistency.
As expected, the performance deteriorates very quickly as soon as the consistency is lost as when comparing to memory, i.e. this task intrinsically requires memory and this is confirmed by comparing the memory and BER maps.
We see again that the best performances are obtained in the center of the consistency region and that a limited resolution analysis gives worse performances in the tongues.
In Figure \ref{fig:figure_5}(d), thanks to the higher resolution we see areas of high performance that are distributed asymmetrically in the tongues. It seems that this asymmetry is related to the dynamic regime and the bifurcations of the system.
Interestingly , it appears that the system can operate with the same level of performance in the center of the area as in the tongues revealed by this study. 
Previous literature on the consistency and performance of a photonic TDRC suggests that the best performance is obtained at the edge of the consistency zone, "at the edge of chaos".
However, the complex structuring of the edge of the consistency region induces a similar structuring on the high performance boundary. Thus, it does not seem sufficient to be "at the edge of" \cite{bertschinger_real-time_2004}. 
Indeed, even with a minimal parametric modification, there is a risk of being on a tongue or between two tongues giving radically different results.
We have also observed that the impact of the modulation parameters is the same on the computer performance as on the consistency. In other words, depending on the values of $\gamma$ and $\phi_{DC}$, the high performance region as well as the tongues shifts and shrinks.

Finally, our numerical simulations also confirm that the tongues leading to high consistency and then good computing performances in the map of parameters, are regions bordered by bifurcations in the laser dynamics. 
More specifically, if the laser diode is chaotic with low consistency property for some set of parameters; when sweeping the parameters to enter the tongues, the laser bifurcates to either time-periodic or quasiperiodic attractors with higher consistency property.
The structured boundary of the consistency region in the parameters map of Figure \ref{fig:figure_2} therefore coincides with a similar structuration of the bifurcation lines.
A full study of the bifurcation lines in a laser with optical feedback and optical injection is however still missing.

\section{Conclusion}
In summary, we have studied a TDRC based on a laser diode and have investigated the impact of the modulation on the consistency and two metrics. 
We have shown that the boundary of the consistency region is not continuous but has a periodic tongue topology. This structuring is only seen when performing a high finesse scanning of the system parameters.
We find that the modulation parameters used, shift and modify the shape of the consistency region and therefore the tongues. Their number and size is directly related to the value of the delay used in the reservoir.
It appears that for the tongues to exist, an adapted modulation depth, a good parametric resolution and an adapted delay are needed.
We show that the RC performance remains of high level in any of these tongues. 
It is therefore important to understand the topology of the consistency boundary to optimise system performance.
It would be interesting to investigate in depth the origin of these tongues through dynamics and complexity analyses and to be able to demonstrate them experimentally.

\textbf{Funding} The Chair in Photonics is supported by Region Grand Est, GDI Simulation, Departement de la Moselle, European Regional Development Fund, CentraleSupelec, Fondation CentraleSupelec, and Eurometropole de Metz.

\textbf{Data Availability Statement} Data underlying the results presented in this paper are
not publicly available at this time but may be obtained from the authors upon
reasonable request.

\textbf{Disclosures} The authors declare no conflicts of interest.

%%%%%%%%%%%%%%%%%%%%%%

\bibliography{biblio.bib}% Produces the bibliography via BibTeX.

\end{document}